# ERLC (Twin LC) and LHC/FCC Based Electron-Proton Colliders

B. Dagli[a], B. Ketenoglu[b,*], S. Sultansoy[a,c]

[a]*TOBB University of Economics and Technology, Ankara, Turkey*
[b]*Department of Engineering Physics, Ankara University, Ankara, Turkey*
[c]*ANAS Institute of Physics, Baku, Azerbaijan*
[*]*Correspondence: bketen@eng.ankara.edu.tr*

**Abstract**

Construction of the ERLC (twin LC) collider tangential to LHC or FCC will give opportunity to realize *ep* collisions at multi-TeV center-of-mass energies. Luminosity estimations show that values well exceeding $10^{34}$ cm$^{-2}$s$^{-1}$ can be achieved for HL-LHC, HE-LHC and FCC based *ep* colliders. Certainly, proposed *ep* colliders have great potential for clarifying QCD basics and new physics search in addition to providing precise PDFs for adequate interpretation of LHC and FCC experimental data.

## 1. Introduction

It is known that construction of future linear colliders tangential to energy frontier hadron colliders will give opportunity to investigate lepton-hadron interactions at highest center-of-mass energies (see review [1] and references therein). This idea was first proposed in mid-1980s for UNK and VLEPP [2, 3]. Then, in 1990s next proposal was the THERA [4]. Among these, the most advanced proposal is the ILC-LHC based QCD-Explorer [5], which has later turned into LHeC [6]. This idea was applied for FCC and SppC in [7] and [8], respectively.

Recently, V. I. Telnov has proposed ERLC (twin LC) scheme to improve ILC luminosity by two orders [9]. Actually, similar schemes were discussed during UNK+VLEPP and THERA studies. However, these options were not included into CDRs. Two years ago, it was mentioned that "as for super high luminosity *ee* collisions at hundreds GeV center-of-mass energies reconsideration of an old idea of energy recovery linear colliders may be useful for ILC" [10]. In our opinion, Telnov's paper will have an essential effect upon decision on construction of the ILC since the luminosity of ERLC is comparable with that of the FCC-*ee*.

In this paper we have considered possible use of the ERLC for HL/HE-LHC and FCC based *ep* colliders. Main parameters of ERLC, HL-LHC, HE-LHC and FCC-*hh* are presented in Section 2. Following section is devoted to evaluation of parameters of corresponding *ep* colliders. In Section 4, we briefly discuss physics search potential of these machines. Our conclusion and recommendations are given in Section 5.

## 2. Parameters of ERLC, LHC and FCC

In this section, we present parameters of ERLC, HL-LHC, HE-LHC and FCC, which are used for estimation of main parameters of *ep* colliders in the following section. Parameters of ERLC and ILC are given in Table 1 (Table 2 of Reference [9]). Table 2 presents nominal parameters for HL-LHC, HE-LHC and FCC [11]. Parameters of proton ring upgraded for ERL60 related *ep* colliders [11] are given in Table 3.

Table 1. Main parameters of ERLC and ILC [10]

| Parameter [unit] | ERLC | ILC |
|---|---|---|
| Beam Energy [GeV] | 125 | 125 |
| N per bunch [$10^{10}$] | 0.5 | 2.0 |
| Norm. emit., $\epsilon_{x,n}$ [μm] | 20 | 10 |
| Norm. emit., $\epsilon_{y,n}$ [μm] | 0.035 | 0.035 |
| $\beta_x$ at IP [cm] | 25 | 1.3 |
| $\beta_y$ at IP [cm] | 0.03 | 0.04 |
| $\sigma_x$ at IP [μm] | 4.5 | 0.73 |
| $\sigma_y$ at IP [nm] | 6.1 | 7.7 |
| Rep. rate, $f$ [Hz] | $2 \times 10^8$ | 6560 |
| Bunch distance [m] | 1.5 | 166 |
| Duty cycle | 1/3 | n/a |

Table 2. Main parameters of HL-LHC, HE-LHC and FCC [11]

| Parameter [unit] | HL-LHC | HE-LHC | FCC |
|---|---|---|---|
| Beam Energy [TeV] | 7 | 13.5 | 50 |
| N per bunch [$10^{11}$] | 2.2 | 2.2 | 1 |
| Norm. emit., $\epsilon_n$ [μm] | 2.5 | 2.5 | 2.2 |
| $\beta_{x,y}$ at IP [m] | 0.15 | 0.45 | 1.1 |
| $\sigma_{x,y}$ at IP [μm] | 7.1 | 9.0 | 6.7 |
| Bunches per beam | 2760 | 2808 | 10400 |
| Rev. rate, $f$ [Hz] | 11245 | 11245 | 3000 |
| Bunch spacing [ns] | 25 | 25 | 25 |
| Bunch length [mm] | 90 | 90 | 80 |

Table 3. LHC/FCC proton beam parameters upgraded for ERL60 based *ep* colliders [11]

| Parameter [unit] | HL-LHC | HE-LHC | FCC |
|---|---|---|---|
| Beam Energy [TeV] | 7 | 13.5 | 50 |
| $\sqrt{s}$ [TeV] | 1.30 | 1.80 | 3.46 |
| N per bunch [$10^{11}$] | 2.2 | 2.5 | 1 |
| Norm. emit., $\epsilon_n$ [μm] | 2 | 2.5 | 2.2 |
| $\beta_{x,y}$ at IP [m] | 0.07 | 0.1 | 0.15 |
| $\sigma_{x,y}$ at IP [μm] | 4.3 | 4.2 | 2.5 |
| Bunches per beam | 2760 | 2808 | 10400 |
| Rev. rate, $f$ [Hz] | 11245 | 11245 | 3000 |
| Bunch spacing [ns] | 25 | 25 | 25 |
| Bunch length [mm] | 90 | 90 | 80 |
| $L_{ep}$, [$10^{34}$ cm$^{-2}$s$^{-1}$] | 0.8 | 1.2 | 1.5 |

## 3. Luminosity of *ep* Colliders

In this section luminosity, disruption and beam-beam tune shift parameters have been calculated using parameters of electron and proton beams given in previous section. Several years ago, the software AloHEP has been developed for estimation of main parameters of linac-ring type *ep* colliders [12, 13]. Recently, AloHEP has been upgraded [14] for all types of colliders (linear, circular and linac-ring) as well as colliding beams (electron, positron, muon, proton and nuclei).

Using the AloHEP software and nominal parameters given in Tables 1 and 2, we have obtained main parameters of the corresponding *ep* colliders given in Table 4. It is seen that ERLC results in more than 2 orders of higher luminosity values compared to ILC. While center-of-mass energy of ERLC based *ep* colliders are higher than that of ERL60 based ones (see Table 3), luminosity values are several times lower. The reason for this is that upgraded LHC and FCC parameters are used for ERL60 based *ep* colliders. Using upgraded LHC and FCC parameters, we have obtained the values given in Table 5 for *ep* colliders under consideration. Here, we have also upgraded ERLC parameters: 5 times lower repetition rate and 5 times higher number of electrons per bunch, since with nominal ERLC bunch distance (1.5 m) only 1/5 of electron bunches collide with proton bunches (bunch distance 7.5 m). As a result, luminosity of ERLC based *ep* colliders exceeds that of ERL60 based ones by more than 2 times.

Table 4. *ep* collider parameters using nominal parameters from Tables 1 and 2

| *p*-beam | *e*-beam | Luminosity [$cm^{-2}s^{-1}$] | $\xi[10^{-4}]$ | D |
|---|---|---|---|---|
| HL-LHC | ILC | $4.6 \times 10^{30}$ | 9.8 | 4.5 |
|  | ERLC | $1.8 \times 10^{33}$ | 2.4 | 4.5 |
| HE-LHC | ILC | $2.9 \times 10^{30}$ | 9.8 | 2.9 |
|  | ERLC | $1.2 \times 10^{33}$ | 2.4 | 2.9 |
| FCC | ILC | $2.3 \times 10^{30}$ | 11 | 2.0 |
|  | ERLC | $9.3 \times 10^{32}$ | 2.8 | 2.0 |

Table 5. *ep* collider parameters for upgraded ERLC and LHC/FCC from Table 3

| Parameter [unit] | HL-LHC | HE-LHC | FCC |
|---|---|---|---|
| $\sqrt{s}$ [TeV] | 1.87 | 2.6 | 5.00 |
| L [$cm^{-2}s^{-1}$] | $2.4 \times 10^{34}$ | $3.0 \times 10^{34}$ | $3.3 \times 10^{34}$ |
| $\sigma_{x,y}$ at IP [μm] | 4.3 | 4.2 | 2.5 |
| Disruption, D | 12 | 15 | 15 |
| Tune Shift, $\xi$ | $1.5 \times 10^{-3}$ | $1.2 \times 10^{-3}$ | $1.4 \times 10^{-3}$ |

Let us mention that luminosity of the ERLC based *ep* colliders can be further improved using dynamic focusing scheme [15]. This scheme should be adopted for vertical plane, since vertical size of the ERLC bunch (Table 1) is much less than vertical size of LHC/FCC bunch (Table 2), while horizontal sizes are close to each other. After all, $L_{ep} \approx 10^{35}$ $cm^{-2}s^{-1}$ seems achievable for all options with reasonable modifications of ERLC and LHC/FCC parameters.

## 4. Physics Potential

Energy frontier *ep* colliders are a must to providing precision PDFs for adequate interpretation of LHC and FCC experimental data. Another goal, in our opinion even more important, is clarifying the QCD basics, especially understanding of confinement. In this context two regions of *x* Bjorken are crucial: small *x* at high $Q^2$ and $x \approx 1$. Highest energy is important for the former and highest luminosity for the latter. Achievable *x* values for ERLC and LHC/FCC based *ep* colliders are presented in Table 6. The last raw of Table 6 is essential for QCD basics, since $Q^2 = 25$ GeV$^2$ certainly corresponds to perturbative QCD, whereas $x<10^{-5}$ means high parton densities.

Table 6. Achievable *x* Bjorken values at ERLC (ERL60) and LHC/FCC based *ep* colliders.

|  | HL-LHC | HE-LHC | FCC |
|---|---|---|---|
| $Q^2 = 1$ GeV$^2$ | $2.9 \times 10^{-7}$ ($6.0 \times 10^{-7}$) | $1.5 \times 10^{-7}$ ($3.1 \times 10^{-7}$) | $4.0 \times 10^{-8}$ ($8.3 \times 10^{-8}$) |
| $Q^2 = 25$ GeV$^2$ | $7.3 \times 10^{-6}$ ($1.5 \times 10^{-5}$) | $3.8 \times 10^{-6}$ ($7.8 \times 10^{-6}$) | $1.0 \times 10^{-6}$ ($2.1 \times 10^{-6}$) |

High luminosity is crucial for investigation of the Higgs boson properties at *ep* colliders. Obviously using of ERLC instead of ERL60 is advantageous since it provides higher center of mass energies and luminosities. Concerning BSM physics, proposed colliders have great potential for a lot of topics such as: leptoquarks, excited electron and neutrino, color octet electron, contact interactions, SUSY, RPV SUSY (especially resonant production of squarks), extended gauge symmetry (especially left-right symmetric models) etc.

## 5. Conclusion

It is shown that construction of ERLC (twin LC) tangential to LHC and FCC will give opportunity to realize multi-TeV center-of-mass energy *ep* colliders with luminosity of order of $10^{34-35}$ cm$^{-2}$s$^{-1}$. Certainly, these colliders will essentially enlarge physics search potential of the LHC and FCC for both the SM and BSM phenomena. In particular, clarification of QCD basics, including confinement, is one of the most important goals.

*Comparison of ERLC and ERL60 based ep colliders.* It is known that construction of ILC-like $e^+e^-$ collider was initially proposed for *ep* collisions at the LHC (see Ref. [1]). However, in order to increase the luminosity, ERL60 came into prominence although the electron beam energy is several times less. We have showed in this study that ERLC is more advantageous for both luminosity and center-of-mass energy aspects. Let us emphasize that LC will provide opportunity to increase the electron beam energy by extending the length of linacs. For instance, 1 TeV $e^+e^-$ will enhance the $\sqrt{S_{ep}}$ by a factor of two, which allows investigation of 4-times smaller values of *x* Bjorken. On the other hand, as shown in Ref. [16] wall plug power of ERL60 is around 200 MW. In order to diminish this value down to 100 MW, ERL beam energy has been decreased from 60 GeV to 50 GeV in the current design [17].

Finally, we believe that systematic studies of accelerator, detector and physics search aspects of ERLC and LHC/FCC based *ep* colliders are necessary for long-term planning in the field of High Energy Physics.

**Acknowledgement**

The authors are grateful to A. N. Akay, U. Kaya and B. B. Oner for useful discussions.